\documentclass[aps,pra,twocolumn,showpacs,10pt,a4paper]{revtex4-1}

\newcommand{\m}[1]{\mathrm{#1}}

\newcommand{\reference}[1]{Ref.~\cite{#1}}
\newcommand{\ket}[1]{|#1\rangle}
\newcommand{\bra}[1]{\langle#1|}
\newcommand{\proj}[1]{|#1\rangle\langle#1|}

\usepackage{graphicx}
\usepackage{color}
\usepackage{cancel}
\usepackage{hyperref}
\usepackage{verbatim}
\usepackage{amsmath}
\usepackage{latexsym}
\usepackage{natbib}
\usepackage{amsmath}
\usepackage{amssymb}
\usepackage{amsthm}
\usepackage{bm}
\usepackage{color}
\usepackage{graphicx}

\newcommand{\nn}{\nonumber}
\newcommand{\kk}{_{\bf{k}}}
\newcommand{\mdag}{^{\dagger}}

\newcommand{\tr}{\mathop{\mathrm{Tr}}\nolimits}
\newcommand{\omg}{\omega}
\newcommand{\eps}{\varepsilon}

\begin{document}

\title{Enhanced thermal stability of the toric code through coupling to a bosonic bath}

\author{Fabio L. Pedrocchi, Adrian Hutter, James R. Wootton, and Daniel Loss}

\address{Department of Physics, University of Basel, Klingelbergstrasse 82, CH-4056 Basel, Switzerland}

\begin{abstract}
We propose and study a model of a quantum memory that features self-correcting properties and a lifetime growing arbitrarily with system size at non-zero temperature. This is achieved by locally coupling a 2D $L \times L$ toric code to a 3D bath of bosons hopping on a cubic lattice. 
When the stabilizer operators of the toric code are coupled to the displacement operator of the bosons, we solve the model exactly via a polaron transformation and show that the energy penalty to create anyons grows linearly with $L$. When the stabilizer operators of the toric code are coupled to the bosonic density operator, we use perturbation theory to show that the energy penalty for anyons scales with $\ln(L)$.  
For a given error model, these energy penalties lead to a lifetime of the stored quantum information growing respectively exponentially and polynomially with $L$. Furthermore, we show how to choose an appropriate coupling scheme in order to hinder the hopping of anyons (and not only their creation) with energy barriers that are of the same order as the anyon creation gaps. We argue that a toric code coupled to a 3D Heisenberg ferromagnet realizes our model in its low-energy sector. Finally, we discuss the delicate issue of the stability of topological order in the presence of perturbations. While we do not derive a rigorous proof of topological order, we present heuristic arguments suggesting that topological order remains intact when perturbative operators acting on the toric code spins are coupled to the bosonic environment.
\end{abstract}


\maketitle

\section{Introduction}
Topologically ordered phases of matter like Kitaev's toric code promise the possibility to store and process quantum information in a manner which is resilient to local imperfections \cite{Kitaev2003,Dennis2002,Bravyi2010,Nussinov2013}. However, a finite gap for the creation of topological defects (called \textit{anyons} in the case of the toric code) is not enough to ensure stability against thermal fluctuations \cite{Nussinov2008,Castelnovo2007,Alicki2007,Alicki2009}. If anyons can be created at a constant energy cost and propagate without any further energy penalty, they will at any non-zero temperature $T$ destroy the stored quantum information in a time which does not increase with the size of the memory. Indeed, it was shown that not only the toric code but a large class of 1-, 2-, and 3-dimensional Hamiltonians suffer from the aforementioned thermal instability of quantum information \cite{Bravyi2009,Haah2010,Yoshida2011,Poulin2013}. This is in contrast to the classical case, where 
 magnetic devices allow the construction of self-correcting hard drives that are stable against both local perturbations and thermal excitations. Proposals  for three-dimensional spin Hamiltonians with local few-spin interactions that do not fall victim to the aforementioned no-go results exist \cite{Bacon2006,Haah2011a,Haah2011b,Michnicki2012}. None of these models is expected to allow for a storage time increasing arbitrarily with system size, while the scaling of the lifetime with temperature may be more favorable than in the bare toric code \cite{Kitaev2003}. A 2D system with a similar behavior has recently been proposed in Ref.~\cite{Benjamin}.

Following a different approach, it has been shown that repulsive long-range interactions between anyons lead to storage times that grow polynomially in $L$ \cite{Beat2010,Chesi2011,Beat2012,Hutter2012}. When the stabilizer operators of the toric code (stabilizers) are resonantly coupled to cavity modes, even a lifetime growing exponentially with $L$ can be achieved \cite{Chesi2011,Hutter2012}. Furthermore, the suppression of anyon diffusion by means of attractive interactions between them has been proposed in Ref.~\cite{Dennis2002} and studied in Ref.~\cite{Hamma2009}. Refs.~\cite{James2011,Cyril2011} studied disorder as a means to hinder quantum propagation of anyons.

In this work, we propose a three-dimensional (3D) model with purely local interactions of bounded strength that presents self-correcting properties. In contrast to the spin-lattice Hamiltonians discussed in Refs.~\cite{Kitaev2003,Bravyi2010,Bravyi2009,Yoshida2011,Bacon2006,Haah2011a,Haah2011b,Michnicki2012} and similar to Ref.~\cite{Hamma2009}, our Hamiltonian involves unbounded bosonic operators.
However, in contrast to Ref.~\cite{Hamma2009} the interaction strengths in our Hamiltonian are bounded while the obtained life-time scalings are more favorable.
We consider a toric code embedded in a 3D reservoir of hopping bosons on a cubic lattice. When the stabilizers are coupled to the bosonic displacement operator, the model is exactly solvable via a polaron transformation. The coupling to the bosons leads to an energy penalty for the anyons that grows linearly with $L$. This is very favorbale since it can lead to a lifetime of the memory that increases exponentially with $L$. This scaling of the lifetime coincides with the four-dimensional toric code \cite{Dennis2002,Alicki2010}, which constitutes so far the only known example of a truly self-correcting quantum memory. We also consider the case when the stabilizers are coupled to the density operator of the bosons, in which case the model is solved with a perturbative second-order Schrieffer-Wolff transformation. We show that the energy penalty for the creation of anyons scales as $\ln(L)$. This scaling of the anyons' gap is in principle sufficient to stabilize the memory and leads to a lifetime increasing polynomially with $L$. 

We present a coupling scheme between stabilizers and bosons that allows to hinder the hopping of anyons, and not only their creation, by energy barriers that are of the same order as the anyon creation gaps, i.e., $O(L)$ or $O(\ln\,L)$. This is useful since imperfections in the initialization process might lead to a finite initial density of anyons.

Furthermore, we argue that a toric code coupled to a 3D Heisenberg ferromagnet in a broken-symmetry state provides a way to realize the proposed Hamiltonian as an effective low-energy theory of a spin-lattice model with bounded operators only.

Finally, we discuss the delicate issue of the stability of topological order in our model. While we do not derive a rigorous proof of topological order, we present heuristic arguments suggesting that topological order remains intact when perturbative operators acting on the toric code spins are coupled to the bosonic environment.

The paper is organized as follows. In Sec. \ref{sec:model} we introduce our model for a toric code embedded in a three-dimensional cubic lattice of hopping bosons. The stabilizer operators are locally coupled to the displacement operator of the bosonic field. In Sec. \ref{sec:main_result} we state that the energetics of the anyon system is accurately described by a Hamiltonian $H_{W}$ with long-range attractive interactions between the stabilizers. This is valid as long as the bosons are in thermal equilibrium with the state of the anyons. We then derive the main result of our work: the energy penalty to slowly create an anyon grows linearly with $L$. We rigorously prove in Sec.~\ref{sec:equivalence} that the energetics of the anyons is indeed described by $H_{W}$. In Sec.~\ref{sec:fast} we consider the fast creation of anyons. We show that the enegy to create an anyon fast is higher than the energy to create it slowly; the energy penalty to create a defect grows in any case linearly with $L$. In Sec.~\ref{sec:densityCoupling} we consider a slightly different model where the stabilizers are locally coupled to the bosonic density operator. This model cannot be treated exactly and we solve it with a perturbative Schrieffer-Wolff transformation. We show that the energy penalty to create an anyon scales as $\ln L$ in this case. In Secs.~\ref{sec:lifetime_linearly} and \ref{app:lifetime} we show that an energy penalty for the anyons scaling with $L$ and $\ln L$ leads to a lifetime of the toric growing respectively exponentially with $L$ and polynomially with $L$.  In Section \ref{sec:ferromagnet} we mention a possible implementation of our model in a Heisenberg ferromagnet. Section \ref{sec:conclusion} contains our final remarks and in particular a discussion of the stability of topological order. Appendix \ref{app:mediated} contains a short review of the Schrieffer-Wolff transformation. In Appendix \ref{app:moments} we calculate all the higher moments ($n\geq2$) of the distribution of energy costs to create an anyon and show that they are all independent of $L$. In Appendix \ref{sec:continuum} we show that the the continuum approximation used in the main text is just a calculational tool that has no influence on the validity of our results.

%%%%%%%%%%%%%%%%%%%%%%%%%%%%%%%%%%%%%%%%%%%%%%%%%%%%%%%%%%%%%%%%%%%%%%%%%%%%%%%%
\section{Coupling to the bosonic displacement operator}\label{sec:model}
\begin{figure}[h]
	\centering
		\includegraphics[width=0.5\textwidth]{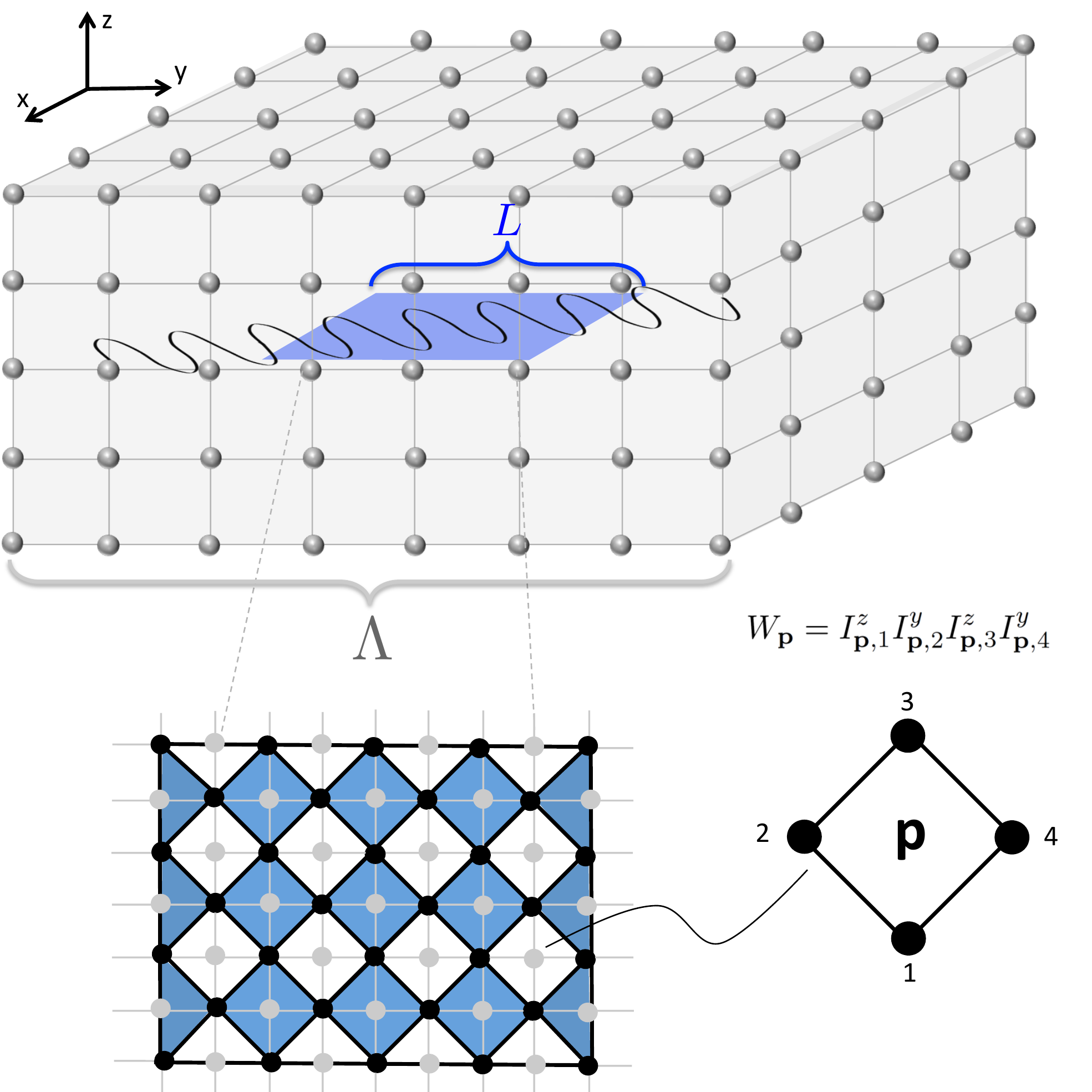}
	\caption{(Color online.) A 2D toric code (blue (dark) area in $xy$-plane) of size $L\times L$ is centered inside  a cubic lattice of size $\Lambda^3$ with $\Lambda>L$. The stabilizers $W_p$ of the toric code locally couple to a system of hopping bosons on a cubic lattice. A long-range attraction between the stabilizers is mediated by the low-energy collective excitations of the bosons.}
	\label{fig:Lattice}
\end{figure}

%%%%%%%%%%%%%%%%%%%%%%%%%%%%%%%%%%%%%%%%%%%%%%%%%%%%%%%%%%%%%%%%%%%%%%%%%%%%%%%%
We present here a model that involves only local interactions of bounded strength in three dimensions. We consider a toric code embedded in a 3D cubic lattice of hopping bosons, see Fig.~\ref{fig:Lattice}.  The stabilizer operators of the toric code are locally coupled to the creation and annihilation operators of the bosons and the total Hamiltonian reads
\begin{equation}\label{eq:boson}
H=H_{\m{b}}+A\sum_{p}W_{p}(a_{p}+a_{p}^{\dagger})\,,
\end{equation}
where the sum runs over the toric code. We denote the linear size of the cubic lattice by $\Lambda$.
Here, the plaquette (stabilizer) operator
$W_{{p}}=I_{{p},1}^{z}I_{{p},2}^{y}I_{{p},3}^{z}I_{{p},4}^{y}$ is the poduct of spins around the square plaquette centered at ${\bf R}_p$, which are defined on a square lattice of linear size $L$ with periodic boundary conditions (we set the lattice constant to unity). To avoid boundary effects we assume $\Lambda>L$.
The 3D vector ${\bf R}_p$ points towards the center of a  plaquette, see Fig.~\ref{fig:Lattice}. Note that this definition of $W_p$ ensures that the blue and white plaquettes are equivalent to the usual toric code star and plaquette operators \cite{Kitaev2003}. The anyon operator $n_{p}$ is defined through $W_{p}=1-2n_{p}$. In other words, when $W_{p}=+1$, the plaquette $p$ carries no anyon and when $W_{p}=-1$, the plaquette $p$ carries an anyon. 
%%%%%%%%%%%%%%%%%%%%%%%%%%%%%%%%%%%%%%%%%%%%%%%%

%%%%%%%%%%%%%%%%%%%%%%%%%%%%%%%%%%%%%%%%%%%%%%%%
The Hamiltonian for the bosons 
\begin{equation}\label{eq:HBOSON}
H_{\m{b}}=\epsilon_{\bf 0}\sum_{i}a_{i}^{\dagger}a_{ i}-t\sum_{\langle i,j\rangle}a_{i}^{\dagger}a_{j}\,,
\end{equation}
 describes bosons hopping on a cubic lattice with hopping amplitude $t$ and on-site chemical potential $\epsilon_{\bf 0}=6t$. Here, $a_{i}^{\dagger}$ creates a boson at site $i$, while $a_{i}$ destroys a boson at site $i$ of the cubic lattice.

Although Hamiltonian (\ref{eq:boson}) is three-dimensional, we point out, for the sake of clarity, that quantum information is stored in the two-dimensional toric code only. As we show below, the presence of the 3D system is necessary to mediate long-range interactions between the stabilizers.

\subsection{Energy of anyon configurations with bosons in thermal equilibrium}\label{sec:main_result}
Here we are interested in the energy penalty to create an anyon. We consider a state with some fixed anyon configuration $\ket{\alpha}$ (i.e., an eigenstate of all operators $W_p$) and with the bosons in thermal equilibrium with respect to that anyon state. 
In other words, the bosons are in the Gibbs state $\rho_\alpha=e^{-\beta H_\alpha}/Z_{\alpha}$ with $Z_{\alpha}=\tr(e^{-\beta H_\alpha})$, $\beta=1/T$, and the bosonic Hamiltonian $H_\alpha = \bra{\alpha}H\ket{\alpha}$ with $H$ defined in Eq.~(\ref{eq:boson}).

In Section \ref{sec:equivalence}, we prove that the energetics of the anyon system is fully described by the diagonal Hamiltonian $H_{W}$, \emph{if the bosons are at each moment in thermal equilibrium $\rho_\alpha$ with respect to the current state $\ket{\alpha}$ of the anyons}.
We have 
\begin{equation}\label{eq:HW}
H_{W}=\sum_{p\neq p'}J_{p,p'}W_{p}W_{p'}\,,
\end{equation}
where $J_{p,p'}$ is a gravitation-like potential between stabilizers, i.e.,
\begin{equation}
J_{p,p'}=-\frac{A^2}{4\pi t\vert {\bf R}_{p}-{\bf R}_{p'} \vert}\,.
\end{equation}
More precisely, in the next subsection we derive the relation (see Eq.~(\ref{eq:equivalence}) below)
\begin{align}\label{eq:energy_anyon}
 \tr\left(\proj{\alpha}\otimes\rho_\alpha\, H\right) = \bra{\alpha}H_W\ket{\alpha} + U_{\m{b}}(\beta)\,,
\end{align}
where $U_{\m{b}}(\beta)$ depends on the temperature $T$ of the bosonic bath but not on the anyon configuration $\ket{\alpha}$.
Since only the first summand depends on $\ket{\alpha}$ and is independent of the temperature of the bosons, the dynamics of the anyon system is described by $H_W$, if the boson system remains in thermal equilibrium with respect to the state of the anyons.
The energy difference between an anyon state $\vert\alpha\rangle$ and another anyon state $\vert\gamma\rangle$ is defined by
\begin{equation}
\Delta E=\text{Tr}(H\vert \alpha\rangle\langle \alpha\vert\otimes\rho_\alpha)-\text{Tr}(H\vert \gamma\rangle\langle \gamma\vert\otimes\rho_\gamma)\,.
\end{equation}
From Eq.~(\ref{eq:energy_anyon}) it directly follows that
\begin{equation}\label{eq:energy_cost}
\Delta E=\bra{\alpha}H_{W}\ket{\alpha} -  \bra{\gamma}H_{W}\ket{\gamma}\,.
\end{equation}
In particular, the energy cost $\Delta E_{0\rightarrow 1}$ to create a single anyon above the anyon-free state $\vert0\rangle$ is
\begin{equation}
\Delta E_{0\rightarrow 1}=\langle 1\vert H_{W}\vert 1\rangle -\langle 0\vert H_{W}\vert 0\rangle\,.
\end{equation}
Note that in the case of periodic boundary conditions, anyons can only be created in pairs. Therefore $\Delta E_{0\rightarrow 1}$ represents a lower bound for the energy gap above the anyonic vacuum, irrespective of the boundary conditions.
In the remaining part of this subsection, we thus study $H_{W}$ and derive how the energy cost $\Delta E_{0\rightarrow1}$ scales with $L$.
This is justified since, as we will show in Sec. \ref{sec:fast}, the energy cost to create an anyon fast enough, such that the thermalization process of the bosons cannot keep pace with the anyon creation, is in fact \emph{higher} than the energy cost $\Delta E_{0\rightarrow1}$.
The Hamiltonian $H_W$ provides thus lower bounds on the energy costs for the creation of an anyon.

Writing $H_{W}$ in terms of anyon operators, $W_{p}=1-2n_{p}$, we obtain
\begin{eqnarray}\label{eq:effective}
H_W=\mu(L)\sum_{p}n_{p}+4\sum_{p\neq p'}J_{{p},{p}'}n_{p}n_{{p}'}+\m{const.}
\end{eqnarray}
The first summand describes a chemical potential for creating an anyon at plaquette $p$, i.e. 
\begin{equation}
\Delta E_{0\rightarrow1}=\mu(L)\,,
\end{equation}
defined by
\begin{equation}
\mu(L) = 4\sum_{p'}(1-\delta_{p,p'})|J_{p,p'}|\,.
\end{equation}
This chemical potential can be evaluated explicitly as
\begin{equation}\label{eq:mainresult}
\mu(L)= \frac{A^2}{\pi\,t}\sum_{p'}\frac{1-\delta_{p,p'}}{\vert{\bf R}_{p'}\vert}\approx \frac{A^2}{\pi\,t}\int_{D_{L/2}}d{\bf R}\,\frac{1}{\vert{\bf R}\vert} = \frac{2A^2}{\,t}\,L\,,
\end{equation}
where we have approximated the square lattice of the toric code by a disk of radius $L/2$ and put the plaquette $p$ and the origin of the coordinate system at the center of the toric code. Note that the continuum approximation used to derive Eq.~(\ref{eq:mainresult}) is a calculational tool to estimate the corresponding sum. Furthermore, in this limit we also let the lattice constant of the surface code go to zero such that a single stabilizer remains coupled to bosonic creation and annihilation operators only at the corresponding site. This approximation is not necessary to obtain the desired behavior since a direct numerical evaluation of the sum shows that it indeed grows linearly with $L$, see Fig.~\ref{fig:numerical_evaluation} in Appendix \ref{sec:continuum}.
Equation (\ref{eq:mainresult}) is a central result of this work; \emph{the chemical potential to create an anyon scales linearly with $L$}. 
In Appendix \ref{app:moments} we also calculate the standard deviation and all higher moments of the distribution of energy costs $\Delta E_{0\rightarrow 1}$~\cite{Poulin_comment}. We show that they are independent of $L$ but increase with temperature $T$, as expected. 
This implies that for any fixed temperature $T$ we can find a size $L$ of the memory such that the distribution of the energy costs is negligible compared to the expected energy cost $\mu(L)$. 

We point out that bosonic operators are not bounded and therefore it is not surprising that the energy cost to create an anyon 
can increase with the size of the system. Qualitatively, our results can be understood as follows. The long-wavelength, low-energy excitations of the bosons mediate a long-range attractive interaction between the stabilizer operators, as as can be seen explicitly in $H_{W}$. Therefore a plaquette feels the presence of all the other plaquettes. In the anyonic vacuum state ($W_{p}=+1$, $\forall\,p$) one needs to overcome the attraction from $L^2-1$ plaquettes in order to create an anyon. Since the interaction between stabilizers decreases with distance, the energy penalty associated to the creation of the anyon scales with $L$ and not with $L^2$.

The second summand in Eq.~(\ref{eq:effective}) describes a gravitation-like interaction between anyons. Since this term helps to keep newly created anyon pairs attached to each other (for temperatures below the interaction strength $\propto A^2/t$), it will have a further beneficial effect on the memory lifetime. On the other hand, this anyon-anyon attraction effectively reduces the anyon chemical potential. However, this reduction is negligible since the anyon density is exponentially suppressed by the first term, see Section \ref{sec:thermally}.

\subsection{Proof of Eq.~(\ref{eq:energy_anyon})}\label{sec:equivalence}
The aim of this subsection is to derive Eq.~(\ref{eq:energy_anyon}).
Let us rewrite Hamiltonian (\ref{eq:boson}) in Fourier space,
\begin{equation}
H=\sum_{\bf q}\epsilon_{\bf q}a_{\bf q}^{\dagger}a_{\bf q}+\frac{A}{\sqrt{N}}\sum_{{p},{\bf q}}W_{{p}}(e^{i{\bf q}\cdot{\bf R}_{p}}a_{\bf q}+\m{h.c.})\,,
\end{equation}
where $a_{\bf q}=\frac{1}{\sqrt{N}}\sum_{i}e^{-i{\bf q}\cdot{\bf R}_{i}}a_{i}$ with $N=\Lambda^3$ the number of lattice sites and $\epsilon_{\bf q}=\epsilon_{\bf 0}-t_{\bf q}$ with $t_{\bf q}=\frac{1}{N}\sum_{\langle ij\rangle}te^{i{\bf q}\cdot({\bf R}_{i}-{\bf R}_{j})}$. Choosing the on-site potential such that $\epsilon_{\bf 0}=t_{\bf 0}=6t$, we obtain the dispersion $\epsilon_{\bf q}=2t\left(3-(\cos(q_x)+\cos(q_y)+\cos(q_z)\right)$.
This Hamiltonian is similar to the independent boson model \cite{Mahan} and thus exactly diagonalizable via 
the unitary polaron transformation
\begin{equation}
\mathcal{S}=-\frac{A}{\sqrt{N}}\sum_{p}W_{p}\sum_{\bf k}\frac{1}{\epsilon_{\bf k}}(a_{\bf k}e^{i{\bf k}\cdot{\bf R}_{p}}-\text{h.c.})\,.
\end{equation}
We have
\begin{eqnarray}
\widetilde{a}_{i}&=&e^{\mathcal{S}}a_{i}e^{-\mathcal{S}}=a_{i}-\frac{A}{N}\sum_{p,\bf q}W_{ p}\frac{1}{\epsilon_{\bf q}}e^{i{\bf q}\cdot({\bf R}_{ i}-{\bf R}_{ p})}\,,\\
\widetilde{a}_{\bf k}&=&a_{\bf k}-\frac{A}{\sqrt{N}}\frac{1}{\epsilon_{\bf k}}\sum_{p}W_{p}e^ {-i{\bf k}\cdot{\bf R}_{p}}\,.
\end{eqnarray}
We thus obtain
\begin{eqnarray}\label{eq:rotated}
\widetilde{H}&=&e^{\mathcal{S}}H e^{-\mathcal{S}}\nn\\
&=&\sum_{\bf q}\epsilon_{\bf q}a_{\bf q}^{\dagger}a_{\bf q}-\frac{A^2}{N}\sum_{{p},{p}'}W_{p}W_{{p}'}\sum_{{\bf q}}
\frac{e^{-i{\bf q}\cdot({\bf R}_{p}-{\bf R}_{p'})}}{\epsilon_{\bf q}}\nonumber\\
&=&\sum_{\bf q}\epsilon_{\bf q}a_{\bf q}^{\dagger}a_{\bf q}+\sum_{p, p' }J_{{p},{p}'}W_{p}W_{{ p}' }\,.
\end{eqnarray}
In order to calculate $J_{p,p'}$, we note that the dominant contributions to $J_{p,p'}$ come from small values of $|{\bf q}|$ (see the integral below) and thus employ a low-${\bf q}$ approximation 
$\epsilon_{\bf q}\approx t{\bf q}^2$. We find
\begin{eqnarray}\label{eq:gravitation}
J_{p,p'}&=&-\frac{A^{2}}{N}\sum_{\bf k}\frac{1}{\epsilon_{\bf k}}e^{i{\bf k}\cdot({\bf R}_{p}-{\bf R}_{{p}'})}\nonumber\\
&=&-\frac{A^2}{(2\pi)^{3}}\int d{\bf k}\frac{1}{\epsilon_{\bf k}}e^{i{\bf k}\cdot({\bf R}_{p}-{\bf R}_{{ p}'})}\nn\\
&\approx&-\frac{A^2}{4\pi t\vert{\bf R}_{p}-{\bf R}_{{p}'}\vert}\,.
\end{eqnarray}
Note that formally $J_{p,p'}$ appears to be divergent for short distances. This, however, is an artefact of the low-${\bf q}$ approximation, which is accurate only for distances $\vert{\bf R}_{p}-{\bf R}_{{p}'}\vert$ sufficiently larger than one lattice constant. We have calculated the integral above for $p=p^{\prime}$ numerically and obtained $J_{p,p}\approx -0.253 A^2/t$.
Since the $p=p'$-terms in $\widetilde{H}$ are irrelevant, we can simply write
\begin{align}\label{eq:Htilde}
\widetilde{H}&=H_{\text{b}}+\sum_{p\neq p' }J_{{p},{p}'}W_{p}W_{{ p}' }+\sum_{p= p' }J_{{p},{p}'}W_{p}W_{{ p}' } \nn\\
&=H_{\text{b}}+H_W+C\, ,
\end{align}
where 
 we used the fact that $W_{p}^{2}=+1$, leading to the irrelevant constant $C$.

Let us define the operator $\mathcal{S}_{\alpha}=\langle \alpha\vert \mathcal{S}\vert \alpha\rangle$.
We now calculate the energy of the state $\proj{\alpha}\otimes \rho_{\alpha}$, where $\vert\alpha\rangle$ is an eigenstate of all $W_{i}$ operators.
Using Eq.~(\ref{eq:Htilde}) and 
\begin{align}
 e^{\mathcal{S}_{\alpha}}e^{-\beta H_\alpha}e^{-\mathcal{S}_{\alpha}} \propto e^{-\beta H_{\m{b}}}
\end{align}
we find
\begin{align}\label{eq:equivalence}
&\text{Tr}(H\vert \alpha\rangle\langle \alpha\vert\otimes\rho_\alpha)\nn\\
&\quad=\text{Tr}(\widetilde{H}e^{\mathcal{S}}\vert \alpha\rangle\langle \alpha\vert\otimes\rho_{\alpha}e^{-\mathcal{S}} )\nonumber\\
&\quad=\text{Tr}(H_{\text{b}}e^{\mathcal{S}_{\alpha}}\rho_{\alpha}e^{-\mathcal{S}_{\alpha}}) +\text{Tr}(H_{W}\,\vert \alpha\rangle\langle \alpha\vert) + C\nonumber\\
&\quad=\text{Tr}(H_{\text{b}}e^{-\beta H_{\m{b}}})/\tr(e^{-\beta H_{\m{b}}}) + \bra{\alpha}H_{W}\ket{\alpha} +C\nonumber\\
&\quad=U_{\m{b}}(\beta) + \bra{\alpha}H_{W}\ket{\alpha}+ C\,,
\end{align}
where $U_{\m{b}}(\beta)$ depends only on the temperature of the bosonic bath but is independent of $\alpha$.  The constant $C$ can be included in $U_{\m{b}}(\beta)$. This completes the proof of Eq.~(\ref{eq:energy_anyon}).

\subsection{Fast creation of an anyon}\label{sec:fast}
In this section, we are interested in the fast creation of an anyon starting from the anyonic vacuum $\ket{0}$, i.e., the state of the toric code with $W_{p}=+1$ for all $p$. We assume that the bosons do not have time to adapt to the creation of an anyon and they remain in their initial equilibrium state $\rho_{0}=e^{-\beta H_{0,\text{b}}}/Z_{0}$ with  $Z_{0}=\text{Tr}(e^{-\beta H_{0,\text{b}}})$ and
\begin{equation}\label{eq:H0}
H_{0,\text{b}}=\bra{0}H\ket{0}=H_{\text{b}}+A\sum_{p}(a_{p}+a_{p}^{\dagger})\,.
\end{equation}
In this case, the chemical potential for an anyon is
\begin{equation}
\Delta E_{0\rightarrow1,\text{fast}}=-2A \langle a_{p}+a_{p}^ {\dagger}\rangle_{0}\,,
\end{equation}
where $\langle O\rangle_{0}=\text{Tr}(O\,e^{-\beta H_{0,\text{b}}})/Z_{0}$.
Defining the operator $\mathcal{S}_0=\bra{0}\mathcal{S}\ket{0}$, we have
\begin{eqnarray}
\widetilde{a}_{ p}&=&e^{\mathcal{S}_0}a_{p}e^{-\mathcal{S}_0}\nonumber\\
&=&a_{p}+\frac{1}{A}\sum_{{p}'}J_{{p},{p}'}\nonumber\\
&=&a_{ p}-\frac{\mu(L)}{4A}-\frac{\vert J_{{p},{p}}\vert}{A}\,.
\end{eqnarray}
We point out again that $|J_{{p},{p}}|$ is finite, see remarks after Eq.~(\ref{eq:gravitation}).
We thus have
\begin{eqnarray}\label{eq:first}
\Delta E_{0\rightarrow1,\text{fast}}
&=&-2A\langle a_{p} + a_{p}^{\dagger}\rangle_0 \nonumber\\
&=&-\frac{2A}{Z_{0}}\text{Tr}(e^{\mathcal{S}_0}e^{-\beta H_{0,\text{b}}}e^{-\mathcal{S}_0}e^{\mathcal{S}_0}(a_{p}+a_{p}^{\dagger})e^{-\mathcal{S}_0}) \nonumber\\
&=&-\frac{2A}{Z_{0}}\text{Tr}(e^{-\beta \widetilde{H}_{0,\text{b}}}(\widetilde{a}_{p}+\widetilde{a}_{p}^{\dagger})) \nonumber\\
&=&-\frac{2A}{Z_{0}}\text{Tr}(e^{-\beta \widetilde{H}_{0,\text{b}}}(a_{p}+a_{p}^{\dagger}))+\mu(L) + 4|J_{{p},{p}}| \nonumber\\
&=&\mu(L) + 4|J_{{p},{p}}|>\mu(L) \,,
\end{eqnarray}
where we used the fact that $\text{Tr}(e^{-\beta \widetilde{H}_{0,\text{b}}}(a_{p}+a_{p}^{\dagger}))=0$ since $\widetilde{H}_{0,\text{b}}=e^{\mathcal{S}_{0}}H_{0,\m{b}}e^{-\mathcal{S}_{0}}=H_{\m{b}}+\m{const.}$

From this calculation we conclude that the energy for the fast creation of an anyon also grows linearly with $L$. 
In fact, it costs more energy to create an anyon fast rather than slowly; this is expected since the bosons do not have time to relax to the new equilibrium configuration.

As noted in Sec.~\ref{sec:main_result}, the origin of the favorable behavior (\ref{eq:first}) resides in the long-range interactions mediated by the low-energy, long-wave length excitations of the bosonic bath. Let us assume that all $W_{p}=+1$. Due to the coupling $A\neq0$ in Eq.~(\ref{eq:boson}), the hopping bosons feel the presence of the plaquettes and the bosonic equilibrium state is populated with bosons such that $\langle a_{p} + a_{p}^{\dagger}\rangle_0\neq 0$. When the size of the toric code increases, more plaquettes are introduced in the system and the population of bosons in the equilibrium state increases, too, i.e., $\langle a_{p} + a_{p}^{\dagger}\rangle_0\sim L$. 

%%%%%%%%%%%%%%%%%%%%%%%%%%%%%%%%%%%%%%%%%%%%%%%%%%%%%%%%%%%%%%%%%%%%%%%%%%%%%%%%%%%%%%%%%%%%%%%%%%%%%%%%%%%%%%%%%%%%%%%%%
\section{Coupling to the bosonic density}\label{sec:densityCoupling}
%%%%%%%%%%%%%%%%%%%%%%%%%%%%%%%%%%%%%%%%%%%%%%%%%%%%%%%%%%%%%%%%%%%%%%%%%%%%%%%%%%%%%%%%%%%%%%%%%%%%%%%%%%%%%%%%%%%%%%%%%
In this section we want to investigate a slightly different model where the stabilizers are locally coupled to the bosonic density $a_{i}^{\dagger}a_{i}$,
\begin{equation}\label{eq:perturbation_z}
H=H_{0}+V=H_{0}+A\sum_{p}W_{p}\,a_{p}^{\dagger}a_{p}\,.
\end{equation}
The main part $H_{0}$ is the Hamiltonian of the hopping bosons, i.e., $H_0=H_{\text{b}}$ and the perturbation $V=A\sum_{p}W_{p}a_{p}^{\dagger}a_{p}$.
In Fourier space the perturbative part in Eq.~(\ref{eq:perturbation_z}) reads
\begin{equation}
V=\frac{A}{N}\sum_{p}W_{p}\sum_{{\bf q},{\bf q}'}e^{i{\bf R}_{p}\cdot({\bf q}-{\bf q}')}a_{\bf q}^{\dagger}a_{{\bf q}'}.
\end{equation}
It is now straightforward to distinguish between the diagonal part $V_{\text{d}}$ and the off-diagonal part $V_{\text{od}}$ of the perturbation, namely
\begin{eqnarray}
V_{\m{d}}&=&\frac{A}{N}\sum_{p}W_{p}\sum_{{\bf q}}a_{\bf q}^{\dagger}a_{{\bf q}}\,,\\
V_{\m{od}}&=&\frac{A}{N}\sum_{p}W_{p}\sum_{{\bf q}\neq{\bf q}'}e^{i{\bf R}_{p}\cdot({\bf q}-{\bf q}')}a_{\bf q}^{\dagger}a_{{\bf q}'}\,.
\end{eqnarray}
Absorbing $V_{\m{d}}$ into the main part of the Hamiltonian, we rewrite
\begin{equation}
H=H'_{0}+V_{\m{od}}\,,
\end{equation}
with
\begin{equation}\label{eq:H0prime}
H'_0=\sum_{{\bf q}}\epsilon_{\bf q}n_{\bf q}+\frac{A}{\Lambda^3}L^2\sum_{\bf q}n_{\bf q}\,,
\end{equation}
where we assumed that the toric code is free of anyons, i.e., $W_{p}=+1$ for all $p$, and we used $N=\Lambda^3$. 

Performing a second-order Schrieffer-Wolff transformation (see App.~\ref{app:mediated})
we obtain the following effective Hamiltonian
\begin{eqnarray}
H_{\text{eff}}&=&-\frac{i}{2}\lim_{\eta\rightarrow0^{+}}\int_{0}^{+\infty}dt\, e^{-\eta t}[V_{\text{od}}(t),V_{\text{od}}]\nonumber\\
&=&\frac{A^2}{2N^2}\sum_{p,p'}W_pW_{p'}\nonumber\\
&&\times\sum_{{\bf q}\neq {\bf q}', {\bf k}\neq {\bf k}'}\frac{e^{i{\bf R}_{p}\cdot({\bf q}-{\bf q}')+{\bf R}_{p'}\cdot({\bf k}-{\bf k}')}}{\epsilon_{\bf q}-\epsilon_{{\bf q}'}}\left[a_{\bf q}^{\dagger}a_{{\bf q}'},a_{{\bf k}}^{\dagger}a_{{\bf k}'}\right]\nonumber\\
&=&\frac{A^2}{2N^2}\sum_{p,p'}W_{p}W_{p'}\sum_{{\bf q}\neq {\bf q}'}\frac{n_{\bf q}-n_{{\bf q}'}}{\epsilon_{\bf q}-\epsilon_{{\bf q}'}}e^{i({\bf q}-{\bf q}')\cdot({\bf R}_{p}-{\bf R}_{p'})}\nonumber\\
&=&\frac{A^2}{2N^2}\sum_{p,p'}W_{p}W_{p'}\sum_{{\bf q}',{\bf k}}\frac{n_{{\bf k}+{\bf q}'}-n_{{\bf q}'}}{\epsilon_{{\bf k}+{\bf q}'}-\epsilon_{{\bf q}'}}e^{i{\bf k}\cdot({\bf R}_{p}-{\bf R}_{p'})}\nonumber\\
&=&-\frac{A^2}{2N^2}\sum_{p,p'}W_{p}W_{p'}\sum_{{\bf q},{\bf k}}\frac{e^{\beta(\epsilon_{{\bf k}+{\bf q}}-\epsilon_{{\bf k} })}}{\epsilon_{{\bf k}+{\bf q}}-\epsilon_{{\bf k}}}n_{{\bf k}+{\bf q}}(n_{{\bf k}}+1)\nonumber\\
&&\hspace{5cm}\times e^{i{\bf q}\cdot({\bf R}_{p}-{\bf R}_{p'})}\nonumber\\
&=&-\frac{A^2}{2N}\sum_{p,p'}W_{p}W_{p'}\sum_{\bf q}\chi({\bf q})e^{i{\bf q}\cdot({\bf R}_{p}-{\bf R}_{p'})}, \nonumber\\
\label{eq:long}\,
\end{eqnarray}
where we introduced  the static `susceptibility' of the bosons 
\begin{equation}\label{eq:susc}
\chi({\bf q})= \frac{1}{N}\sum_{{\bf k}}\frac{e^{\beta(\epsilon_{{\bf k}+{\bf q}}-\epsilon_{{\bf k} })}}{\epsilon_{{\bf k}+{\bf q}}-\epsilon_{{\bf k}}}n_{{\bf k}+{\bf q}}(n_{{\bf k}}+1).
\end{equation}
Following the approach of Ref.~\cite{Kawasaki} assuming that $\beta\epsilon_{{\bf q}+{\bf k}},\beta\epsilon_{\bf q},\beta(\epsilon_{{\bf k}+{\bf q}}-\epsilon_{\bf k})\ll1$, we have that 
\begin{equation}\label{eq:zz}
\chi({\bf q})=\frac{T}{8 t^2}\frac{1}{\vert{\bf q}\vert}\,\,\,\,\m{for}\,\,\,\vert{\bf q}\vert\rightarrow0\,.
\end{equation}
The effective Hamiltonian then becomes
\begin{align}\label{eq:Heff}
 H_{\text{eff}} &= -\frac{A^2T}{16t^2}\sum_{p,p'}W_{p}W_{p'}\frac{1}{N}\sum_{\bf q}\frac{1}{\vert{\bf q}\vert}e^{i{\bf q}\cdot({\bf R}_{p}-{\bf R}_{p'})} \nn\\
&= -\frac{A^2T}{16t^2}\sum_{p,p'}W_{p}W_{p'}\frac{1}{(2\pi)^3}\int d{\bf q}\frac{1}{\vert{\bf q}\vert}e^{i{\bf q}\cdot({\bf R}_{p}-{\bf R}_{p'})} \nn\\
&= -\frac{A^2T}{32\pi^2t^2}\sum_{p,p'}W_{p}W_{p'}\frac{1}{|{\bf R}_{p}-{\bf R}_{p'}|^2}\ .
\end{align}
The interaction strength between the stabilizers mediated by the bosons decays now with the square of the inverse distance ($1/R^{2}$) rather than with the inverse distance ($1/R$), as in the previous section.
Furthermore, the coupling strength is proportional to temperature.

The Schrieffer-Wolff transformation we performed is nothing but a unitary transformation $e^{-\mathcal{S}}$ (similar to the polaron transformation) up to second order in the small parameter $A/t$. Therefore, the same line of reasoning as in Sec.~\ref{sec:equivalence} applies and the energetics of the anyons is fully described by $H_{\text{eff}}$. 
In other words, the energy difference $\Delta E$ between two states $\proj{\alpha}\otimes\rho_{\alpha}$ and $\proj{\gamma}\otimes\rho_{\gamma}$ is
\begin{equation}
\Delta E\approx \text{Tr}(H_{\text{eff}}\vert\alpha\rangle\langle\alpha\vert)-\text{Tr}(H_{\text{eff}}\vert\gamma\rangle\langle\gamma\vert)\,,
\end{equation}
where the sign $\approx$ means that the effective Hamiltonian is calculated up to second order only.

From Eq.~(\ref{eq:Heff}), we finally find a chemical potential for the anyons that grows now logarithmically with $L$,
\begin{equation}
\mu(L)\sim \frac{A^2T}{t^2}\ln(L/2)\,,
\end{equation}
where we used
\begin{equation}
\int_{D_{L/2}}\, d^{2}R\, \frac{1}{R^2}\sim \ln (L/2)\,.
\end{equation}
%%%%%%%%%%%%%%%%%%%%%%%%%%%%%%%%%%%%%%%%%%%%%%%%%%%%%%%%%%%%%%%%%%%%%%%%
\section{Thermally Stable Quantum Memory}\label{sec:thermally}
%%%%%%%%%%%%%%%%%%%%%%%%%%%%%%%%%%%%%%%%%%%%%%%%%%%%%%%%%%%%%%%%%%%%%%%%
As we have demonstrated in the previous sections, coupling the toric code stabilizers to a 3D bath of hopping bosons has a very beneficial effect: 
the energy penalty to create an anyon grows with $L$ if we couple to the bosonic displacement operator and with $\ln(L)$ if we couple to the bosonic density.
Here we show that a toric code with an anyon chemical potential growing linearly or logarithmically with $L$ has respectively a lifetime growing exponentially or polynomially with $L$.
The physical picture behind this is that it takes longer and longer for the anyons to reach their thermodynamic equilibrium state with increasing values of $L$~\cite{Beat2010,Chesi2011,Beat2012,Hutter2012}.

\subsection{Anyon chemical potential  linear in $L$}\label{sec:lifetime_linearly}
A chemical potential for anyons in the toric code that grows linearly with $L$ leads to a quantum information storage time that grows exponentially with $L$ and $\beta$, where $\beta=1/T$ is the inverse temperature of a bath weakly coupled to the memory. This follows from Sec.~8 in \reference{Chesi2010}. Assuming that the interaction with the thermal bath can be described by the Davies equation and that the thermal state is a fixed point of the Lindblad operators, the authors of \cite{Chesi2010} proved that the lifetime of the memory $\tau$ scales as $\tau=O(e^{\beta\mu}/L^2)$, where $\mu$ is the anyons' chemical potential. Here, we present alternative arguments leading to the same conclusion: when the anyons' chemical potential is $\mu(L)$, the lifetime of the toric code is at least $\tau= O(e^{\beta \mu(L)}/L^2)$.
In Sec.~\ref{app:lifetime} we will show that if $\mu(L)$ grows slow enough, this lower bound is no longer tight and the actual lifetime-scaling is more favorable.

Let us try to understand in more detail the decoherence process of the memory in contact with a simple model of a bath. We assume that the bath supports single-spin processes in which an energy $\omega$ is transfered from the anyon system to the bath with rate $\gamma(\omega)$ and that $\gamma(0)\neq0$ \cite{ohmic_comment}. Let $\delta(N)$ denote the average cost to create an anyon pair if there are already $N$ pairs present. The gravitational interaction will lead to $\delta(N\geq 1)<\delta(0)=2\mu(L)-A^2/(4\pi t)$. However, below we show that this reduction will not lead to a finite self-consistent number of anyon pairs and that in fact we will have $\delta(N\geq 1)\approx\delta(0)$ in the relevant regime.

Since the presence of only two anyons diffusing across the memory leads to an uncorrectable logical error in times of order $L^2/\gamma(0)$ \cite{Beat2010}, we need to show that the time for the creation of two nearby anyons that are not directly annihilated increases exponentially with system size. Whenever a new pair of anyons is created, their total hopping rate is given by $6\gamma(0)$ \cite{hopping_comment} such that the probability that one of the two anyons ever moves before the pair gets annihilated is $6\gamma(0)/[\gamma(\delta(0))+6\gamma(0)]$. 
Since $\gamma(\delta(0))=\exp(\beta\delta(0))\gamma(-\delta(0))$ (which follows from the detailed balance condition) and the code consists of $L^2$ physical spins, we conclude that the total rate for creation of anyon pairs that do not directly get annihilated is given by 
\begin{equation}
L^2\gamma(-\delta(0))\frac{6\gamma(0)}{\gamma(\delta(0))+6\gamma(0)} \leq 6L^2 e^{-\beta\delta(0)}\gamma(0)\ .
\end{equation}
The time needed to create such a pair is thus of order $\exp(\beta\delta(0))/L^2\gamma(0)$. 
In conclusion, we found a lower bound for the quantum memory storage time that increases exponentially with $\delta(0)$. Since $\delta(0)$ is linear in $L$, the lifetime increases exponentially with $L$.

Assume that there are already $N$ anyon \textit{pairs} present. We want to determine the average (averaged over all possible positions of the existing anyons) energy cost $\delta(N)$ to create a new pair. From the point of view of one of the two newly created anyons, we assume that the existing $2N$ anyons are uniformly distributed over all $L^2-2$ remaining positions. The averaged interaction between one of the newly created anyons and each existing one is thus
\begin{eqnarray}
&&\frac{1}{L^2-2}\left(4\sum_{{p}\neq { 0}}|J_{{p},{ 0}}|+A^2/(4\pi t)\right) \nonumber\\
&&= -\frac{1}{L^2-2}\left(2\mu(L)-A^2/(4\pi t)\right)\ ,
\end{eqnarray}
where we have subtracted the energy $-A^2/(4\pi t)$ due to attraction with the other anyon of the same pair. Indeed, we are only interested in the attraction energy due to anyons which are already present before the creation of the pair. The total energy $\delta(N)$ to create the new pair is thus given by
\begin{eqnarray}
\delta(N) &=& \delta(0) -\frac{4N}{L^2-2}\left(2\mu(L)-A^2/(4\pi t)\right) \nonumber\\
&=& \delta(0)\left(1-\frac{4N}{L^2-2}\right)\ ,\label{eq:deltaM}
\end{eqnarray}
where $\delta(0)=2\mu(L)-A^2/(4\pi t)$.

The mean-field energy of $N$ anyon pairs is thus
\begin{equation}\label{eq:Emf}
E_{\m{mf}}(N) = \sum_{i=0}^{N-1}\delta(i) = \delta(0)N\frac{L^2-2N}{L^2-2}\, .
\end{equation}
The symmetry $N\leftrightarrow L^2/2-N$ is reminiscent of the fact that the energy of $H_W$ in Eq.~\eqref{eq:HW} can be minimized by either all stabilizers having a $+1$ eigenvalue (no anyons present) or a $-1$ eigenvalue (memory full of anyons). 
The energetic gap between the sector in which there are almost no anyons and the sector in which the memory is full of anyons is of order $\delta(0)L^2=O(L^3)$, so transitions between these two sectors happen on time-scales much longer than the time before the stored quantum information is lost.
Consequently, each sector may serve as a thermally stable quantum memory, but at each moment in time we can only use one of the two. Without loss of generality, we consider the case where the sector with (almost) no anyons present is used for quantum information storage.

From Eq.~(\ref{eq:deltaM}) we have that $\delta(N)=\delta(0)(1-2 n)$, where $n$ denotes the density of anyons.
As there can only be zero or one anyon at each position, we obtain the self-consistent equation for the mean-field anyon density \textit{in equilibrium}
\begin{equation}\label{eq:selfConsistent}
n_{\m{mf}} = [\exp\left(\beta\delta(0)(1-2n_{\m{mf}})\right)+1]^{-1}\ .
\end{equation}
If the left-hand side of this equation is smaller/larger than the right-hand side, the anyon density will tend to increase/decrease. If $n_{\m{mf}}$ solves this equation, so does $1-n_{\m{mf}}$.
One self-consistent density is $n_{\m{mf}}=\frac{1}{2}$.
The stability of this density depends on the temperature of the bath. For $\beta\delta(0)<2$ we have a unique self-consistent density $n_{\m{mf}}=\frac{1}{2}$ and this density is also stable.
For $\beta\delta(0)>2$ the density $\frac{1}{2}$ becomes unstable and two new stable self-consistent densities $n^*$ and $1-n^*$ emerge (let $n^*$ denote the smaller of the two).
The system of gravitationally interacting anyons therefore shows a phase transition and spontaneous breaking of the anyon anyon-hole symmetry at a critical temperature $\delta(0)/2$, which is of order $\frac{A^2}{t}L$.
For the purpose of quantum information storage, we are clearly interested in temperatures below this critical temperature. 

Adding the usual toric code Hamiltonian \cite{Kitaev2003} $H_{\m{toric}}=-\frac{\Delta}{2}\sum_{p}W_{p}$ to Eq.~\eqref{eq:boson} explicitly breaks the symmetry between anyons and anyon holes and will lead to an additional summand $2N\Delta$ in Eq.~(\ref{eq:Emf}). However, the modification of the self-consistent densities $n^*$, $1-n^*$, and $\frac{1}{2}$ through this new term becomes vanishing for large $L$, as $\Delta$ does unlike $\delta(0)$ not grow with $L$. 

Let us consider the self-consistent solution $n^{*}$. We want to show that $n^{*}$ is exponentially suppressed with $L$ and consequently that the number of anyons itself goes to zero in the thermodynamic limit. After straightforward algebra, one can show that $n=2e^{-\beta\delta(0)}<1/2$ with $\beta\delta(0)e^{-\beta\delta(0)}<\frac{\log(2)}{4}$ (note that this condition is readily satisfied since $\delta(0)$ grows linearly with $L$) satisfies
\begin{equation}\label{eq:nAssumption}
[\exp(\beta\delta(0)(1-2n))+1]^{-1}<n \ ,
\end{equation}
and therefore $n>n^{*}$. Since $n$ is by definition exponentially suppressed with $L$ and $n^{*}<n$ we finally conclude that the self-consistent solution $n^{*}$ of Eq.~(\ref{eq:selfConsistent}) goes exponentially to zero with $L$. A direct consequence of this is that the equilibrium number of anyons $n^{*}L^2$ also vanishes exponentially with $L$ and will generally be much smaller than the minimal positive value $2$. Hence the anyon number will fluctuate between $0$ and small even integers, such that $\delta(N)\approx\delta(0)$ from Eq.~(\ref{eq:deltaM}).

\subsection{Anyon chemical potential  logarithmic in $L$}\label{app:lifetime}
Here we show that a chemical potential growing logarithmically with $L$ leads to a lifetime of the memory growing polynomially with $L$.

By the same line of reasoning as in Sec.~\ref{sec:lifetime_linearly}, modifications to the anyon chemical potential due to inter-anyonic interactions are negligible.
Let us thus study a simple model in which anyons have a constant energy cost $\mu$ independent of the number of anyons which are already present.
Ref.~\cite{Chesi2010} predicts in this scenario a lifetime that scales at least with $\exp(2\beta\mu)/L^2$ \cite{pair_comment}. Employing the same simple bath model as in the previous paragraph, let us probe the tightness of this bound. As remarked in Sec.~\ref{sec:lifetime_linearly}, it takes a time of order $t_{1}=\exp(2\beta\mu)/(L^2\gamma(0))$ to create an anyon pair that does not immediately annihilate but performs at least one hopping. One such separating pair creates an uncorrectable logical error in times of order $\sim L^2/\gamma(0)$. We ignore here dimensionless $O(1)$ factors which depend on the precise definition of the memory lifetime and on the classical algorithm employed to perform error correction. Thus if we are in the regime $\mu>2T\ln L$, the quantum information will get destroyed by the first separating pair, which takes a time of order $t_1$ such that the bound in Ref.~\cite{Chesi2010} is tight. 

However, consider now the opposite regime $\mu<2T\ln L$. In this regime, further anyons will be created before the two anyons of the first separating pair have time to diffuse across a distance of order $L$. The lifetime of the memory is then given by the time it takes the anyons to diffuse across the average inter-pair distance, which is when error correction will inevitably break down. After a time $t$, the density of anyons will be of order $t/(t_1L^2) = \gamma(0)t\times\exp(-2\beta\mu)$, taking the possibility for immediate annihilation into account, and existing anyons will have diffused across a distance $\sim\sqrt{\gamma(0)t}$, as the diffusion constant for anyons is essentially given by $\gamma(0)$ \cite{Beat2010}. Consequently, after a time $\sim \exp(\beta\mu)/\gamma(0)$ existing anyons will have diffused across the current inter-pair distance, which thus constitutes the lifetime of the memory. Notably, in this case the bound from Ref.~\cite{Chesi2010} is no longer tight, as $\exp(\beta\mu)>\exp(2\beta\mu)/L^2$ in the assumed regime.

To summarize, if anyons can be created at a constant energy cost $\mu$ and the quantum memory is in contact with a bath that supports processes which have an energy cost $\omega$ with a rate $\gamma(-\omega)$ and fulfills the detailed balance condition, error correction will break down after a time of order
\begin{eqnarray}
&&\left.\begin{array}{cl} \exp(2\beta\mu)/L^2 \gamma(0), & \mbox{if }\mu\geq2T\ln L\\ \exp(\beta\mu)/\gamma(0), & \mbox{if }\mu\leq2T\ln L \end{array}\right\}\nonumber\\
&=&\max\left\lbrace\exp(2\beta\mu)/L^2 , \exp(\beta\mu)\right\rbrace /\gamma(0).
\end{eqnarray} 

Now let us assume that $\mu=\mu(L)=cT\ln L$, which is what we have obtained in Sec.~\ref{sec:densityCoupling} when coupling the stabilizers to the boson density. Then we obtain a lifetime scaling as $\max\left\lbrace L^{2c-2} , L^c\right\rbrace/\gamma(0)$, i.e., polynomially growing for any $c>0$ with a change in the scaling behavior, depending on whether $c$ is greater or smaller than $2$. 
However, recall that in our case $c\sim A^2/t^2\ll1$, such that the lifetime grows only modestly with $L$.

We note that for bath models as employed in Refs.~\cite{Beat2010,Beat2012,Hutter2012}, we have $\gamma(0)\propto T$, so our estimate for the lifetime contains an implicit temperature-dependence, even though the explicit temperature dependence stemming from the Boltzmann factor drops out.

%%%%%%%%%%%%%%%%%%%%%%%%%%%%%%%%%%%%%%%%%%%%%%%%%%%%%%%%%%%%%%%%%%%%%%%%%%%%%%%%%%%%%%%%
\section{Hindering of anyon hopping}\label{app:hindering}
%%%%%%%%%%%%%%%%%%%%%%%%%%%%%%%%%%%%%%%%%%%%%%%%%%%%%%%%%%%%%%%%%%%%%%%%%%%%%%%%%%%%%%%%
The lifetime of the memory that we discussed above does not apply if the initial state of the system has anyons already present. Suppose that errors occur during preparation of the initial state, creating a finite density of anyons. If these errors are sufficiently sparse, it will be possible for error correction to recover the initial state. It is the job of the Hamiltonian to preserve this error correctability until the desired time of readout. The coupling of the quantum memory to the hopping bosons will energetically favour the annihilation of anyons on neighboring plaquettes, undoing some of the errors. However, we can expect that a finite density of pairs will have been non-neighbouring, and so will remain. These only need to diffuse a constant distance to make correction ambiguous, which leads to a constant lifetime for the memory. To prevent this we can split the plaquettes into two types. `Strongly coupled' plaquettes are coupled to the hopping bosons with a strength $A_s$. `Weakly coupled' plaquettes have a strength $A_w < A_s$. These are chosen such that any sequence of single- or local two-spin errors that move an anyon from one weakly coupled plaquette to another must move it via a strongly coupled plaquette. Example patterns are given below. The chemical potential for the plaquettes will change from the form in Eq.~\eqref{eq:mainresult}, giving different values $\mu_s (L)$ and $\mu_w (L)$ for the two types of plaquette. Performing the summation (as described in the following subsection) shows that the factor $A^2$ in Eq.~\eqref{eq:mainresult} becomes $A_s\bar A$ for $\mu_s (L)$ and $A_w\bar A$ for $\mu_w (L)$ ($\bar A$ being a weighted average of $A_s$ and $A_w$). The energy barrier required for anyon movement is therefore of order $(1-A_w/A_s)\mu_s(L)$, which increases linearly with system size. The resulting suppression of diffusion leads to a lifetime that increases exponentially with system size, even when the initial state has a finite density of anyons.

\begin{figure}[h]
	\centering
		\includegraphics[width=0.45\textwidth]{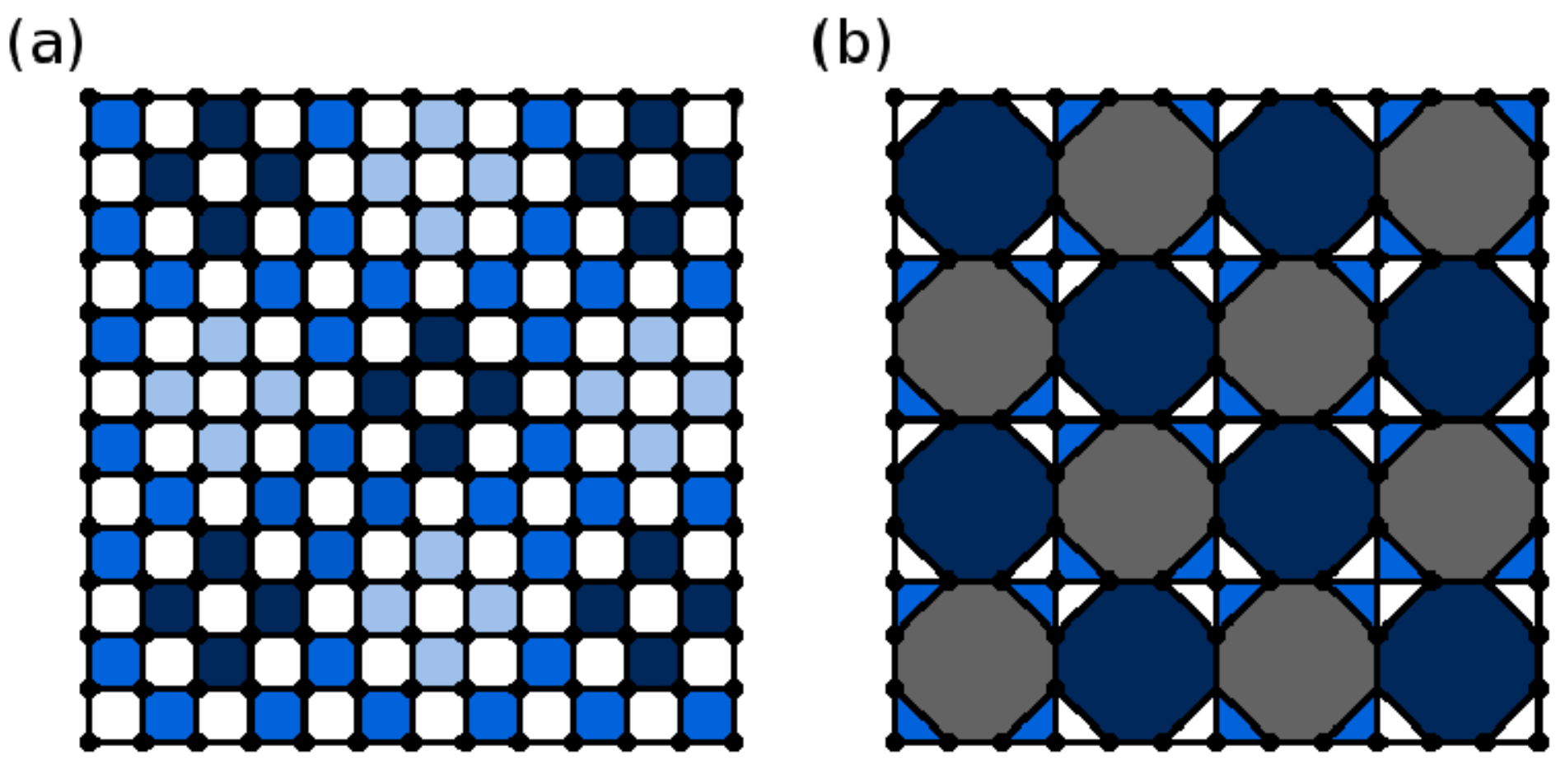}
	\caption{(Color online.) Two tilings of plaquettes are shown on which the code may be defined. Spins are located on vertices. (a) The square tiling, as usually employed for the toric code. $s$-plaquettes are shown in dark blue (black), blue (grey), or light blue (light grey), $p$-plaquettes are shown in white. (b) An alternative tiling, with alternating triangular and octagonal plaquettes. $s$-plaquettes are shown in dark blue (black) and blue (grey), $p$-plaquettes are shown in white and grey (light grey).}
	\label{fig:Lattices}
\end{figure}

It may come as a surprise that associating some stabilizers with a lower energy penalty has a beneficial effect on the memory. However, note that the weakly coupled plaquettes allow energy to be dissipated from the anyons to the bath by hopping of an anyon from a strongly to a weakly coupled plaquette. On the other hand, if the chemical potential is independent of the anyon position, as in Eq.~\ref{eq:mainresult}, this is only possible through annihilation of anyons.

\subsection{An example pattern for strongly and weakly coupled plaquettes}
We will now look at the concepts proposed above in greater detail and find specific examples for patterns of strongly and weakly coupled plaquettes.

In the toric code model there are two types of anyon, $e$ and $m$, which reside on two kinds of plaquette, $s$ and $p$, respectively. Note that, when the code is defined with spins on the edges of the lattices, these correspond to the stars and plaquettes, respectively.

Consider a spin in the square tiling of Fig.~\ref{fig:Lattices} (a), shared by two $s$-plaquettes to the top-left and bottom-right and two $p$-plaquettes to the top-right and bottom-left. The application of a Pauli $I^z$ to such a spin will affect the $e$ anyon occupations of the two $s$-plaquettes. If both were initially empty, an anyon pair will be created. If both initially held an anyon, this pair will be annihilated. If only one held an anyon, it will be moved to the other plaquette. The application of a Pauli $I^y$ has the same effect for the $m$ anyons of the $p$-plaquettes. For spins where the positions of $s$- and $p$-plaquettes are exchanged, the roles of $I^z$ and $I^y$ are also exchanged. No operation exists that can move an anyon from an $s$-plaquette to a $p$-plaquette, or vice-versa.

Creation, movement and annilation of anyons are therefore achieved by Pauli operations. Using single spin operations, creation of a pair will always lead to the anyons occupying neighboring plaquettes (where neighoring means that they share exactly one spin). Similarly, single spin operations can only move anyons from one plaquette to a neigboring one, or annihilate anyons on neighboring plaquettes. Since we assume that the system-bath coupling supports only single spin errors, it is exactly these processes that we consider during thermalization. However, it should be remembered that two-spin perturbations may also be present in the Hamiltonian. Local two-spin errors should therefore also be considered, which can create, annihilate and transport anyons on next-to-neighboring plaquettes.

With this in mind, we wish to split both $s$- and $p$-plaquettes into two groups, one of which will be strongly to the hopping bosons with a coupling $A_s$ and the other of which will be weakly coupled with a strength $A_w<A_s$. This will give the plaquettes of the former a higher chemical potential than those of the latter, with an energy difference that increases linearly with system size.

The pattern of strongly and weakly coupled plaquettes should be chosen such that anyons become trapped within the latter, which will occur if two conditions are satisfied. Firstly, any anyons initially on strongly coupled plaquettes should quickly move into a nearby weakly coupled plaquette. Secondly, it should not be possible for anyons to be moved from one weakly coupled plaquette (or a small cluster of weakly coupled plaquettes) to another by a sequence of either single- or two-spin operations without passing through a strongly coupled plaquette.

The first condition can be met if anyons on strongly coupled plaquettes cannot be moved over large distances by a sequence of either single or two spin operations without either moving through a weakly coupled plaquette, or through a strongly coupled plaquette that neigbors a weakly coupled one. The latter is relevant because it will ensure that the distance an anyon can move before decaying into a weakly coupled plaquette is exponentially suppressed.

Both conditions are satisfied by the pattern shown in Fig.~\ref{fig:Lattices} (a). Here, weakly coupled $s$-plaquettes are shown in dark blue. Strongly coupled $s$-plaquettes that neighbor weakly coupled $s$-plaquettes are shown in blue, and those that do not are shown in light blue. Regions of strongly coupled plaquettes that do not neighbor weakly coupled plaquettes are separated from each other by a width of three spins. Sequences of one- and two-spin operations therefore cannot move anyons in one such region to  another without going via strongly coupled plaquettes that do neighbor weakly coupled plaquettes, which will almost certainly result in the anyon decaying into the neighboring weakly coupled plaquettes. Similarly, regions of weakly coupled plaquettes are separated by the same width, preventing movement between them without going via strongly coupled plaquettes.

The initial movement of anyons on strongly coupled plaquettes to nearby weakly coupled plaquettes may cause ambiguity for error correction if the error rate during initialization is too high. Even so, for sufficiently low error rates this movement will have no effect on correctability. Once the movement is complete, the exponential suppression of diffusion will then ensure that the correctability of the errors is preserved for a time exponential with the system size, since such an exponentially long timescale will be required for the anyons to climb out of the weakly coupled plaquettes.

We will now demonstrate that the difference in chemical potentials between strongly and weakly coupled plaquettes leads to the energy barrier required to suppress diffusion. To determine the chemical potential of an arbitrary plaquette ${p}$ (which is either $s$- or $p$-type), the following sum over all plaquettes must be performed
\begin{equation}
\mu_{{p}}(L)=\frac{M^2}{2\pi R} A_{{p}} \sum_{{p}'}\!^{'} A_{{p}'} \frac{1}{\vert{p}-{p}'\vert}\ ,
\end{equation}
where the prime in $\sum\!^{'}$ means that ${p}'\neq {p}$. Here $A_{{p}'}$ denotes the coupling of plaquette ${{p}'}$ which will be $A_s$ or $A_w$ depending on whether this plaquette is weakly or strongly coupled, respectively. By numerically performing the summation we find that, in the $L \rightarrow \infty$ limit, it takes the form
\begin{equation}
\sum_{{p}'}\!^{'}  A_{{p}'}\frac{1}{\vert{\bf R}_{p}-{\bf R}_{p'}\vert} = \frac{3 A_s + A_w}{4} \, c  L\ ,
\end{equation}
where $c=4\ln(1+\sqrt{2})\simeq3.53$ is defined via
\begin{align}
 \int_{[-L/2,L/2]^2}\frac{dxdy}{\sqrt{x^2+y^2}}=cL\,.
\end{align}
The linear combination of $A_s$ and $A_w$ is a weighted average $\bar A = (3 A_s + A_w)/4$, which arises from the fact that there are three times as many strongly coupled plaquettes as weakly coupled plaquettes. The chemical potentials for weakly and strongly coupled plaquettes are then
\begin{equation}
\mu_s(L) = \frac{c A_s \bar A M^2}{2\pi R} \cdot L\ , \,\,\,\,\, \mu_w(L) = \frac{c A_w \bar A M^2}{2\pi R} \cdot L\ .
\end{equation}
Clearly, $\mu_s(L)-\mu_w(L) = O(L)$, giving the required energy barrier.

\subsection{Alternative tiling with four-body coupling}

A pattern of strongly and weakly coupled plaquettes, stable against single-spin errors, is shown for an alternative tiling in Fig.~\ref{fig:Lattices} (b). Strongly (weakly) coupled $s$-plaquettes are shown in blue (dark blue) and strongly (weakly) coupled $p$-plaquettes are shown in white (grey). For this tiling it is still true that $e$ anyons can only be created and moved between neighboring $s$-plaquettes, and $m$ anyons between neighboring $p$-plaquettes. Note that all strongly coupled plaquettes in this tiling are triangular. The $W_{{p}}$ for these will therefore be three-body operators, making the code-hopping boson coupling only a four-body term. On the other hand, weakly coupled plaquettes are octagons with eight-body $W_{{p}}$ and nine-body terms required for the code-hopping boson coupling. Since these many-body terms will most likely be generated by perturbative methods, with a higher number of spins in a term generated by higher orders of perturbation theory, the difference in coupling strengths will arise naturally.

Due to the practical difficultly in generating many-body terms, we can consider not coupling the octagonal plaquettes to the hopping bosons. Only the four-body terms required to couple the triangles are then needed, which should be easier to implement than the five-body terms required for the square tiling. Despite the fact that only a fraction of the plaquettes are coupled to hopping bosons, the memory is still stable against thermal errors. This is because any single spin error must still create at least one anyon on, or move anyons through, energetically penalized triangular plaquettes. The energy barrier that increases linearly with system size is therefore still intact, and ensures that anyon creation and diffusion are exponentially suppressed.

Unfortunately, stability against local Hamiltonian perturbations does not remain strong without the coupling of octagons. Without an energy penalty, two-body perturbations are free to create and move anyons between next-to-neighboring octagonal plaquettes. This avoids the energy barrier and so leads to uncorrectable errors in a constant time. However, it is possible to avoid this by carefully considering what types of perturbation are present, and then designing the $W_{{p}}$ such that they are unable to perform such hopping processes. For example, let us use $W_{{p}} = I_{{p},1}^{x}I_{{p},2}^{y}I_{{p},3}^{z}$ for triangular $s$-plaquettes. Here spin $1$ is that shared with the neighboring triangular $s$-plaquette and the numbering proceeds clockwise. Let us also use $W_{{p}} = I_{{p},1}^{z}I_{{p},2}^{y}I_{{p},3}^{x}$ for triangular $p$-plaquettes with corresponding numbering. No nearest neighbor isotropic perturbation of the form $I^{\alpha}_i I^{\alpha}_j$, for $\alpha \in \{x,y,z\}$, commutes with all of these operators. This means such perturbations will be suppressed by the energy barrier and will not be able to move anyons between octagonal plaquettes. If only perturbations of this form are present in the system, the memory will remain stable.

\section{Ferromagnet as bosonic bath}\label{sec:ferromagnet}

In this section, we would like to point out a physical system where bosonic modes (as discussed in the previous sections) naturally occur as a lowest order approximation.
Indeed, the Hamiltonians (\ref{eq:boson}) and (\ref{eq:perturbation_z}) are closely related to the Hamiltonians describing a toric code embedded in a 3D Heisenberg ferromagnet (FM) in a broken-symmetry state at finite temperature. More explicitly, let us consider the following Hamiltonian
\begin{equation}\label{eq:ferrox}
H'=H_{F}+A\sqrt{2/S}\sum_{p}W_{p}S_{p}^{x}\,,
\end{equation}
where 
\begin{equation}\label{eq:fm}
H_{\m{F}}=-J\sum_{\langle i,j\rangle}{\bf S}_{i}\cdot{\bf S}_{j}+h_{z}\sum_{i}S_{i}^{z}
\end{equation}
 is the Hamiltonian of a 3D Heisenberg ferromagnet (FM) of linear size $\Lambda\gg L$,  where  $J>0$ 
is the exchange coupling constant and  the sum is restricted to nearest-neighbor lattice sites. The FM is assumed to be below the Curie temperature and the spins ordered  along the $z$-direction. We can now perform a Holstein-Primakoff transformation~\cite{Nolting}
\begin{equation}\label{eq:HP}
S_{i}^{z}=-S+{\hat n}_{i}\,,\,\,\,\,S_{i}^{-}=a_{i}^{\dagger}\sqrt{2S-{\hat n}_{i}} \,,\,\,\, S_{i}^{+}=(S_{i}^{-})^{\dagger},
\end{equation}
in the formal limit ${\hat n}_{i}\ll 2S$, where ${\hat n}_{i}=a_{i}^{\dagger}a_{i}$~\cite{Nolting}. It is then straightforward to show that the low-energy sector of Hamiltonian (\ref{eq:ferrox}) is equivalent to Hamiltonian (\ref{eq:boson}). Following the same reasoning, we conclude that the Hamiltonian
\begin{equation}
H''=H_{F}+A\sum_{p}W_{p}S_{p}^{z}
\end{equation}
is in its low-energy sector equivalent to Hamiltonian (\ref{eq:perturbation_z}). 

However, since all operators in Hamiltonians $H'$ and $H''$ are bounded, it is clear that the energy penalty for flipping a toric code spin very fast cannot grow without bounds as a function of $L$ \cite{Poulin_comment}.
Still, it seems reasonable to expect that for adiabatic noise sources, that drag the FM along while flipping a spin, the response of the FM resembles the one of the bosonic bath studied in this work, since the stabilizers are coupled via the susceptibility of the FM.
It is thus reasonable to assume that the toric code might be protected against such adiabatic noise sources when embedded in the FM.
Note that the question of how to engineer five-spin interactions, as required for Hamiltonians $H'$ and $H''$, remains open.

%%%%%%%%%%%%%%%%%%%%%%%%%%%%%%%%%%%%%%%%%%%%%%%%%%%%%%%%%%%%%%%%%%%%%%%%%%%%%%%%
\section{Conclusions and discussion}\label{sec:conclusion}
%%%%%%%%%%%%%%%%%%%%%%%%%%%%%%%%%%%%%%%%%%%%%%%%%%%%%%%%%%%%%%%%%%%%%%%%%%%%%%%%
In this paper we have introduced a 3D-model with purely local, bounded-strength interactions in three dimensions that is self-correcting at finite temperatures. Our model is exactly solvable and consists of a toric code locally coupled to a system of hopping bosons on a cubic lattice. The stabilizer operators are locally coupled to the displacement operator of the bosons and a long-range attractive interaction between stabilizer operators is mediated by the low-energy collective excitations of the bosonic system. This leads to a chemical potential for the anyons growing linearly with $L$ and can be used to stabilize the quantum memory against thermal fluctuations. For a given error model, a chemical potential of the anyons that grows linearly with $L$ leads to a lifetime of the quantum memory increasing exponentially with $L$. When the stabilizers are coupled to the bosonic density, a chemical potential growing only with $\ln L$ is derived. We show that such a chemical potential is enough to stabilize the memory whose lifetime increases polynomially with $L$. 

If the degeneracy of the highly entangled states which form the code subspace is not robust against local perturbations, uncontrolled splitting of this degeneracy induced by local imperfections would lead to dephasing of the logical qubit. 
It was already argued in Ref.~\cite{Kitaev2003} and rigorously proved in Ref.~\cite{Bravyi2010} that for the standard toric code Hamiltonian \cite{Kitaev2003}, perturbations which are weak enough (compared with the anyon creation gap), time-independent, and local (or exponentially decaying) lead to a lifting of the groundstate degeneracy that is exponentially small in $L$. 
Since our Hamiltonian is not gapped and involves unbounded operators, the result of Ref.~\cite{Bravyi2010} do not apply. 
While we consider a rigorous treatment of this issue to be beyond the scope of the present work, which focuses on stability against thermal errors rather than perturbations, we briefly present arguments suggesting that robustness to local perturbations is valid in our model. 

As pointed out in Ref.~\cite{BondersonPRB}, in any real solid the degrees of freedom that do not directly constitute the ``memory'' (spins of the toric code) represent a gapless environment to which the memory couples.
This situation is not addressed by studies of perturbations which act entirely within the Hilbert space of the memory, as is the case in Ref.~\cite{Bravyi2010}. The issue of accidental couplings to gapless modes is therefore by no means unique to our quantum memory proposal and will be present in any physical implementation of a quantum memory.
In Ref.~\cite{BondersonPRB} the authors discuss topological phases coupled to a gapless environment, and find that in some cases (``strong quasi-topological phases'') the topological properties, including the exponentially suppressed groundstate splitting, survive this coupling.
Such strong quasi-topological phases, including the toric code coupled to a gapless environment (such as accoustic phonons or photons), thus constitute the strongest form of a quantum memory one could hope for in nature -- except for the fact that they are not thermally stable.
Our memory \emph{is} thermally stable and in the following we present heuristic arguments that in our system couplings to the gapless modes may  not pose a threat to the topological order either. 

Recall that engineered couplings of strength $A$ (see Hamiltonian (\ref{eq:boson})) between the stabilizer operators and the bosonic modes lead to an anyon creation gap of the order $O(\frac{A^2}{t}L)$. 
Now consider accidental couplings of the form $\eps I^x_i(a_i+a_i\mdag)$, where $I^x_i$ is a bit-flip that acts on a physical qubit of the toric code.
In second-order perturbation theory, the coupling to the bosonic field leads to terms of the form $\frac{\eps A}{t}\sum_{i\neq j}I_{i}^{x}W_{j}/\vert{\bf R}_{i}-{\bf R}_{j}\vert$ and $\frac{\eps^2}{t}\sum_{i\neq j}I_{i}^{x}I^{x}_{j}/\vert{\bf R}_{i}-{\bf R}_{j}\vert$.
Summing over all plaquettes the former terms take the form $O(\frac{\eps A}{t}L)I^x_i$; the condition that these perturbations are sufficiently weak compared to the anyon creation gap simply translates into the requirement that $\eps$ is sufficiently small compared to $A$, i.e., that the accidental couplings are sufficiently weak compared to the engineered ones. 
The second-order terms describing interactions between bit-flips are weaker and will have support only on two small regions, which for most pairs $i$ and $j$ are well-separated. This does not allow anyons to hop non-locally, as would be required to distinguish the ground states. Despite their non-local form, these perturbations are therefore still similar in effect to local perturbations.
We thus believe that our Hamiltonian is robust against this type of perturbations and splitting of the ground-state degeneracy is well-suppressed with $L$. However, a rigorous proof remains a very interesting open question.

%%%%%%%%%%%%%%%%%%%%%%%%%%%%%%%%%%%%%%%%%%%%%%%%%%%%%%%%%%%%%%%%%%%%%%%%
\section{Acknowledgements}
%%%%%%%%%%%%%%%%%%%%%%%%%%%%%%%%%%%%%%%%%%%%%%%%%%%%%%%%%%%%%%%%%%%%%%%%
We would like to thank 
%D.~DiVincenzo, F.~Hassler, B.~Terhal, and 
D.~Poulin for helpful discussions, and L.~Trifunovic for pointing out the connection to the independent boson model. This work was supported by the Swiss NSF, NCCR Nanoscience, and NCCR QSIT.
%%%%%%%%%%%%%%%%%%%%%%%%%%%%%%%%%%%%%%%%%%%%%%%%%%%%%%%%%%%%%%%%%%%%%%%%%%%%%%%%%%%%%%%%%%%%%%%%%%%%%%%%%%%%%
\appendix

\section{Schrieffer-Wolff transformation}\label{app:mediated} 
For the sake of completeness, we present in this appendix the derivation of the second order  Schrieffer-Wolff transformation (for a general discussion see  \cite{BravyiSW2011}). We start from
\begin{equation}
H=H_0+V\,,
\end{equation}
where we identify $H_{0}$ as the main part and $V$ as a small perturbation. We decompose the spectrum $\sigma(H_{0})$ of $H_{0}$ into a high-energy set of eigenvalues $M_{Q}$ and a low-energy set of eigenvalues $M_{P}$ such that $\sigma(H_0)=M_{P}\cup M_{Q}$, $M_{P}\cap M_{Q}=\emptyset$, and there is a gap separating the eigenvalues in $M_{P}$ and $M_{Q}$. We define the operators $P$ and $Q=1-P$ respectively as the projectors onto the low energy subspace $\mathcal{M}_{P}$ and onto the high-energy subspace $\mathcal{M}_{Q}$ corresponding to set of eigenvalues $M_{P}$ and $M_{Q}$. The perturbation $V$ can then be decomposed into a diagonal part $V_{\m{d}}$ and an off-diagonal part $V_{\m{od}}$
\begin{eqnarray}
V_{\m{d}}&=&PVP+QVQ\,,\\
V_{\m{od}}&=&PVQ+QVP\,.
\end{eqnarray}
The effective Hamiltonian is given by a Schrieffer-Wolff transformation such that the transformed Hamiltonian $H_{\m{eff}}=e^{S}He^{-S}$ is block-diagonal, i.e., $PH_{\m{eff}}Q=QH_{\m{eff}}P=0$.
Up to second order in $V$ the effective Hamiltonian reads \cite{Bernd2008,BravyiSW2011}
\begin{equation}
H_{\m{eff}}^{(2)}=H_{0}+V_{\m{d}}+U=H'_{0}+U\,,
\end{equation}
where we define $H'_{0}=H_{0}+V_{\m{d}}$ and
\begin{equation}\label{eq:U}
U=-\frac{i}{2}\lim\limits_{\eta\rightarrow0^{+}}\int_{0}^{\infty}\m{d}t\,e^{-\eta t}\left[V_{\m{od}}(t),V_{\m{od}}\right]\ ,
\end{equation}
where $V_{\m{od}}(t)=e^{iH'_{0}t}V_{\m{od}}e^{-iH'_{0}t}$ is given in the Heisenberg representation.

\section{Standard deviation and higher moments of the distribution of energy costs}\label{app:moments}

Let us now calculate the standard deviation of the distribution of the energy costs to create an anyon. 
For simplicity, we consider the case of fast changes, where all relevant thermal expectation values are given by $\langle\ldots\rangle_0$,
which denotes thermal averages with respect to the original thermal state of the bosons.

The standard deviation is given by
\begin{align}
 \sigma_{\text{fast}} = \sqrt{\langle(2A (a_{ p}+a_{p}^{\dagger}))^{2}\rangle_0-\langle 2A(a_{p}+a_{p}^{\dagger})\rangle_0^{2}}\,.
\end{align}
We first consider
\begin{align}\label{eq:standard}
&\langle(2A(a_{p}+a_{p}^{\dagger}))^{2}\rangle_0 \nn\\
&\quad=\frac{4A^{2}}{Z_{0}}{\text{Tr}(e^{-\beta H_{0,\text{b}}}(a_{p}+a_{p}^{\dagger})^{2})}\nn\\
&\quad=\frac{4A^{2}}{Z_{0}}{\text{Tr}(e^{-S}e^{-\beta \widetilde{H}_{0,\text{b}}}(\widetilde{a}_{p}^{2}+(\widetilde{a}_{p}^{\dagger})^{2}+1+2\widetilde{a}_{p}^{\dagger}\widetilde{a}_{p})e^{S})}\,.\nn\\
\end{align}
We have
\begin{eqnarray}\label{eq:a}
\widetilde{a}_{p}^{2}&=&a_{p}^{2}-2\frac{\mu(L)+4|J_{pp}|}{4A}a_{p}+((\mu(L)+4|J_{p,p}|)/4A)^2\nn\\
(\widetilde{a}_{p}^{\dagger})^{2}&=&(a_{p}^{\dagger})^{2}-2\frac{\mu(L)+4|J_{p,p}|}{4A}a_{p}^{\dagger}+((\mu(L)+4|J_{p,p}|)/4A)^2\nn\\
\widetilde{a}_{p}^{\dagger}\widetilde{a}_{p}&=&a_{p}^{\dagger}a_{p}-\frac{\mu(L)+4|J_{p,p}|}{4A}(a_{p}+a_{p}\mdag) 
\nn\\ &&\quad 
+((\mu(L)+4|J_{p,p}|)/4A)^2\ .
\end{eqnarray}
By inserting (\ref{eq:a}) into (\ref{eq:standard}), and using the fact that
\begin{eqnarray}
\text{Tr}(e^{-S}e^{-\beta \widetilde{H}_{0}} a_{p}e^{S})&=&\text{Tr}(e^{-S}e^{-\beta \widetilde{H}_{0}} a_{p}^{\dagger}e^{S})\nn\\
&=&0\,,
\end{eqnarray}
we obtain
\begin{align}
 \langle(2A(a_{p}+a_{p}^{\dagger}))^{2}\rangle_0 
&=(\mu(L)+4|J_{p,p}|)^2 + 4A^2 + \nn\\&\quad +8A^2\frac{\tr(e^{-\beta H_{\text{bos}}} a_{p}\mdag a_{p})}{\tr(e^{-\beta H_{\text{b}}})}\,.
\end{align}
Furthermore, we have shown in Eq.~\eqref{eq:first} that
\begin{align}
 -2A\langle a_{p} + a_{p}\mdag\rangle_0 = \mu(L) + 4|J_{p,p}|\ .
\end{align}
In conclusion,
\begin{align}\label{eq:sigmafast}
 \sigma_{\m{fast}} 
&= 2A \sqrt{1 + 2\frac{1}{N}\sum\kk\frac{1}{e^{\beta\omg\kk}-1}} \nn\\
&\simeq 2A \sqrt{1 + 2\frac{4\pi}{(2\pi)^3}\int_0^\infty\m{d}k\frac{k^2}{e^{\beta Dk^2}-1}} \nn\\
&= 2A \sqrt{1 + \frac{\zeta(3/2)}{4(\pi\beta D)^{3/2}}}\ .
\end{align}
We see that the standard deviation is of order $A$, slowly increases with temperature, and, crucially, is independent of $L$, such that $\frac{\sigma_{\m{fast}}}{\mu(L)}\sim\frac{t}{A L}$ becomes negligible for large $L$.

Let us now calculate the higher moments of the distribution.
In order to simplify our notation, we define $X_{p}= -2A(a_{p}+a_{p}^{\dagger})$, such that the expected energy cost is $\Delta E_{0\rightarrow1,\text{fast}} = \langle X_{p}\rangle_0$.
We define the $n$-th moment of the distribution to be
\begin{equation}
 C_n = \left\langle(X_{p} - \langle X_{p}\rangle_0)^n\right\rangle_0^{1/n}\ .
\end{equation}
We find
\begin{eqnarray}
 C_n^n 
&=& \left\langle\sum_{k=0}^n\binom{n}{k}X_{p}^{n-k}(-1)^k\langle X_{p}\rangle_0^k\right\rangle_0 \nn\\
&=& \sum_{k=0}^n(-1)^k\binom{n}{k}\left\langle X_{p}^{n-k}\right\rangle_0\langle X_{p}\rangle_0^k\ .
\end{eqnarray}
Now in order to evaluate these averages we write 
\begin{equation}\label{eq:avg}
 \left\langle X_{p}^{m}\right\rangle_0 = \left\langle e^{-S}\widetilde{X}_{p}^{m}e^S\right\rangle_0 = \left\langle \widetilde{X}_{p}^{m}\right\rangle_\text{b}
\end{equation}
where $\langle\ldots\rangle_\text{b}$ denotes thermal averages w.r.t.\ $H_\text{b}$ and $\widetilde{X}_{p}=e^SX_{p}e^{-S}=X_{p}+\mu(L)+4|J_{p,p}|$.
For the second equality in Eq.~\eqref{eq:avg} we have used the fact that $\widetilde{H}_0=H_\text{b}+\m{const}$.
Then, using Wick's Theorem and the fact that $\langle (a_{p}+a_{p}^{\dagger})^{2k+1}\rangle_{\text{b}}=0$,
\begin{eqnarray}
\left\langle X_{p}^{m}\right\rangle_{0} &= &\sum_{k=0}^m\binom{m}{k}(\mu(L)+4|J_{p,p}|)^{m-k}\langle X_{p}^k\rangle_\text{b} \nn\\
&=& \sum_{k=0}^{\lfloor m/2\rfloor}\binom{m}{2k}(\mu(L)+4|J_{p,p}|)^{m-2k}\langle X_{p}^{2k}\rangle_\text{b} \nn\\
&= &\sum_{k=0}^{\lfloor m/2\rfloor}\binom{m}{2k}\frac{(2k)!}{2^kk!}(\mu(L)+4|J_{p,p}|)^{m-2k}\langle X_{p}^{2}\rangle^k_\text{b}\ .\nonumber\\
\end{eqnarray}
For the last equality, we have used that the number of possible contractions is $(2k-1)\times(2k-3)\ldots3\times1=\frac{(2k)!}{2^kk!}$. 
As simplest case, we have $\langle X_{p}\rangle_{0}=\mu(L)+4|J_{p,p}|$. In conclusion, we find
\begin{widetext}
\begin{eqnarray}
 C_n^n &=& \sum_{k=0}^n(-1)^k\binom{n}{k}\sum_{r=0}^{\lfloor (n-k)/2\rfloor}\binom{n-k}{2r}\frac{(2r)!}{2^rr!}(\mu(L)+4|J_{p,p}|)^{n-k-2r}\langle X_{p}^{2}\rangle^r_\text{b}(\mu(L)+4|J_{p,p}|)^k \nn\\
&= &n!(\mu(L)+4|J_{p,p}|)^n\sum_{k=0}^n\sum_{r=0}^{\lfloor (n-k)/2\rfloor}\frac{(-1)^k}{k!r!(n-k-2r)!} \left(\frac{\langle X_{p}^{2}\rangle_\text{b}}{2(\mu(L)+4|J_{p,p}|)^2}\right)^r\ .
\end{eqnarray}
\end{widetext}
This sum can be evaluated by use of the identity
\begin{align}
 \sum_{k=0}^n\sum_{r=0}^{\lfloor (n-k)/2\rfloor}\frac{(-1)^k}{k!r!(n-k-2r)!} \xi^r = \begin{cases} \frac{\xi^{n/2}}{(n/2)!}\ , &\text{if $n$ is even} \\ 0\ , &\text{if $n$ is odd} \end{cases}\ .
\end{align}
We thus obtain, for $n$ even,
\begin{align}
 C_n = \left(\frac{n!}{(n/2)!}\right)^{1/n} \sqrt{\langle X_{p}^{2}\rangle_\text{b}/2}\ . 
\end{align}
Furthermore,
\begin{align}
 \sqrt{\langle X_{p}^{2}\rangle_\text{b}} &= 2A\left(1+2\frac{1}{N}\sum\kk\langle a\kk\mdag a\kk\rangle_\text{b}\right) \nn\\
&=2A \sqrt{1 + \frac{\zeta(3/2)}{4(\pi\beta t)^{3/2}}}\ ,
\end{align}
see Eq.~\eqref{eq:sigmafast}.

Our final result is thus
\begin{align}
 C_n = \sqrt{2}A\left(\frac{n!}{(n/2)!}\right)^{1/n} \sqrt{1 + \frac{\zeta(3/2)}{4(\pi\beta t)^{3/2}}}
\end{align}
for $n$ even, and $0$ otherwise. For $n=2$ we retrieve (\ref{eq:sigmafast}) for the the standard deviation.
For larger $n$, recall that $\left(\frac{n!}{(n/2)!}\right)^{1/n}\approx\sqrt{2n/e}$,
such that
\begin{align}
 C_n \approx 2A\sqrt{n/e} \sqrt{1 + \frac{\zeta(3/2)}{4(\pi\beta t)^{3/2}}}\ .
\end{align}
In conclusion, all the higher moments grow like $O((T/t)^{3/4})$ with temperature but are independent of $L$.

\section{Continuum approximation}\label{sec:continuum}
Here we numerically evaluate the sum $\sum_{ p}\frac{1}{\vert {\bf R}_{p}\vert}$ and show that the continuum approximation is just a convenient mathematical tool that allows to analytically evaluate the behavior of the sum as function of $L$.

\begin{figure}[h]
	\centering
		\includegraphics[width=0.45\textwidth]{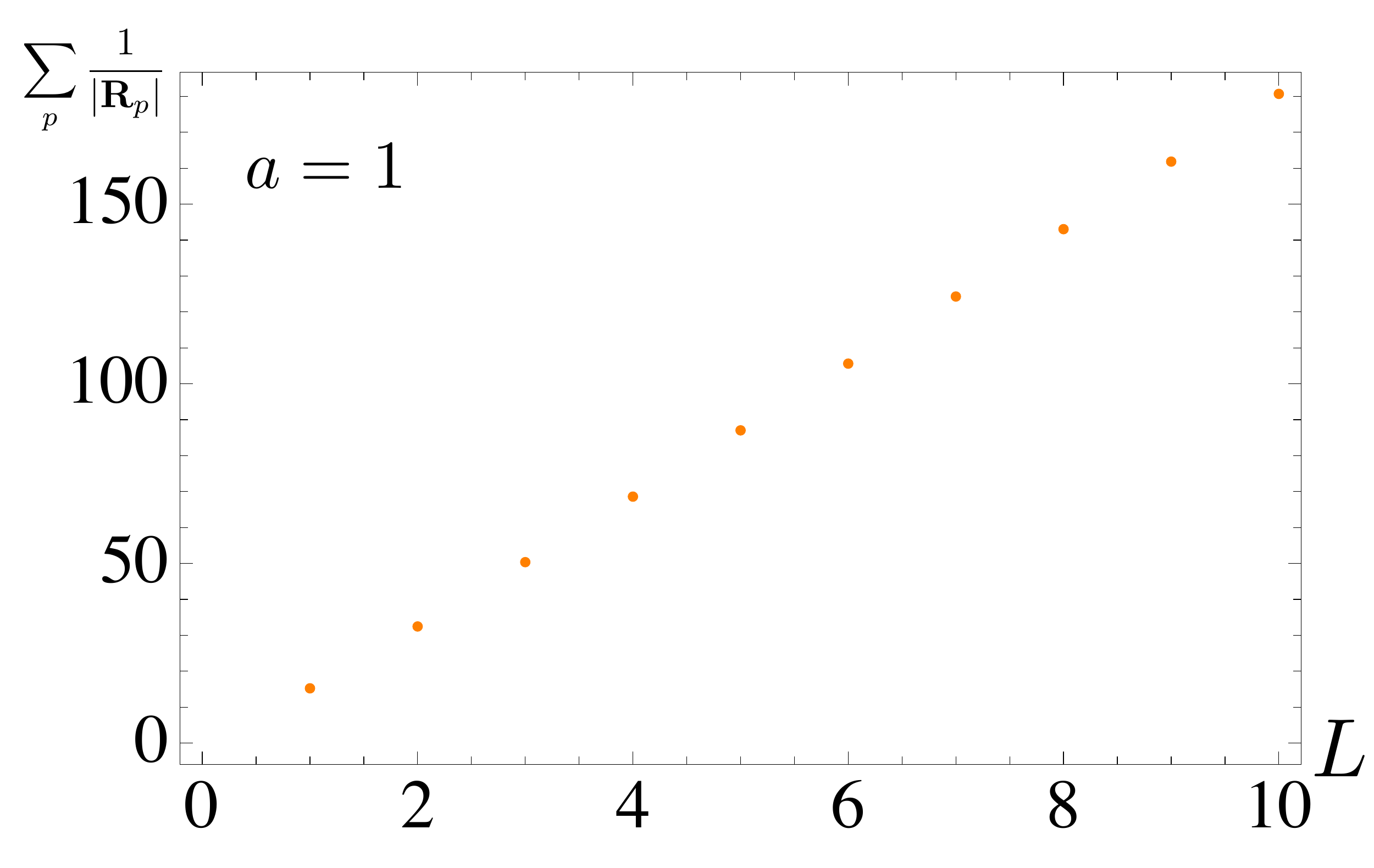}
	\caption{Numerical evaluation of the sum $\sum_{ p}\frac{1}{\vert {\bf R}_{p}\vert}$ as function of $L$ for a lattice constant $a=1$. The sum increases linearly with $L$, in agreement with the continuum approximation calculation.}
	\label{fig:numerical_evaluation}
\end{figure}

In Fig.~(\ref{fig:numerical_evaluation}) we plot the sum $\sum_{ p}\frac{1}{\vert{\bf R}_{p}\vert}$ as function of $L$. Here we choose $a=1$ for the lattice constant. The linear behavior is in agreement with the continuum approximation calculation. The other sums appearing in this work can similarly be evaluated numerically and the results agree with the continuum  approximation. As mentioned in the main text, we point out again that in the continuum approximation we let the lattice constant $a$ of the surface code go formally to zero such that a single stabilizer remains coupled to a bosonic creation an annihilation operators  at the corresponding site.

\begin{thebibliography}{100}
\bibitem{hopping_comment}
Strictly speaking, the energy cost for hopping is greater than zero since it increases the potential energy in the gravitational potential. However, we approximate this energy cost by zero for simplicity, neglecting the beneficial effect of the anyon attraction and obtaining a lower bound on the actual lifetime.
\end{thebibliography}
\end{document}